\documentstyle[12pt,epsfig]{article}

\def\lab#1      {\hbox{\small #1} }
\newcommand{\be}{\begin{eqnarray}}
\newcommand{\ee}{\end{eqnarray}}
\newcommand{\ben}{\begin{eqnarray*}}
\newcommand{\een}{\end{eqnarray*}}
\newcommand{\la}{\langle}
\newcommand{\ra}{\rangle}
\newcommand{\half}{\frac{1}{2}}
\newcommand{\pe}{\rightarrow}

\def\mb#1         {\mbox{\boldmath $#1$}}
\def\diffn#1	  {\Delta^{-}_{#1}}

\begin{document}

\begin{titlepage}

\begin{tabbing}
\` {\sl hep-lat/0108004} \\
    \\
\` LSUHE No. 381-2001 \\
\` August, 2001 \\
\end{tabbing}

\vspace*{0.1in}

\begin{center}

{\large\bf 
Center vortices on SU(2) lattices
\\}
\vspace*{.5in}
A. Alexandru$^1$
and Richard W. Haymaker$^2$\\
\vspace*{.2in}
{\small\sl Department of Physics and Astronomy, Louisiana State
University,\\ Baton Rouge, Louisiana 70803-4001, U.S.A.
\\
\vspace{3ex}
e--mail addresses:      $^1$alexan@rouge.phys.lsu.edu, 
                  $^2$haymaker@rouge.phys.lsu.edu}

\end{center}
\vspace{0.2in}
\begin{center}
\begin{minipage}{5in}
\begin{center}
{\bf Abstract}
\end{center}

\end{minipage}
\end{center}

We show that gauge invariant definitions of thin, thick 
and hybrid center vortices,  defined by Kovacs and Tomboulis on $SO(3) \times Z(2)$ 
configurations, can also be defined in $SU(2)$.   We make this connection  using the freedom of 
choosing a particular $SU(2)$ representative of $SO(3)$. 
We further show that in another representative the Tomboulis $\sigma - \eta$ thin vortices
are P (projection) vortices.  The projection approximation corresponds to dropping the
perimeter factor of a Wilson loop after appropriate gauge fixing.   We present results for 
static quark potentials based on these vortex counters and compare projection vortex counters with
gauge invariant ones on the same configuration.

\vfill

\noindent

PACS indices:

11.15.Ha, 

11.30.Ly. 

\end{titlepage}

\section{Introduction}

Given the possibility that SU(N) center vortices occur abundantly 
in the vaccuum of an $SU(N)$  lattice gauge theory then a simple 
mechanism emerges that may  account for color confinement.
A Wilson loop that links the  core but lies in the asymptotic tail of the vortex
would pick up a factor of $Z(N)$ compared to a configuration without this singular gauge transformation.  
If the cost of action of inserting this vortex is minimal then averages could be
disordered by a mechanism that is proportional to the area of the loop.  This has led
to many recent efforts to understand vortices 
in the vacuum of QCD\cite{dfgo,kt1,elrt,dp}. 

Efforts to confirm this picture on the lattice remain problematic.  Kovacs and Tomboulis\cite{kt2}  
were able to confirm that a center vortex in SU(2) could survive in the continuum limit at vanishing
cost of action.   In this example the center vortex is topologically trapped by twisted boundary
conditions and there is no doubt about its presence.  But it is not simple to establish 
the presence of a center vortex in an SU(N) lattice configuration that is not trapped.   

In addition to these structures,  there are thin vortices associated with negative 
plaquettes costing action proportional to the area of the vortex sheet.  
Contrary to the above case these are suppressed in the continuum limit.   

Some time ago, Tomboulis\cite{t}
developed a formalism that discriminates between these two structures.  In this alternative partition 
function, the links, $U(b)$, are elements of $SU(2)$  but the 
action is invariant under sign flips of the links.
Consequently, the two equivalent link values correspond to two equivalent
{\em representatives} of $SO(3)$.  The $Z(2)$ factor group is manifest on new
independent variables $\sigma(p)$ living on plaquettes.  The bookkeeping arising from the change in 
representatives is carried by $Z(2)$ valued  dependent variables $\eta(p) = \;sign \;\lab{tr} [ U(\partial p)]$.

Recently we studied the configuration space of this formulation further in order to do simulations in these
variables\cite{ah1}.  We found an ergodic algorithm for simulations on lattices with periodic and twisted boundary 
conditions.   In Section 2 of this paper 
we find a mapping between these variables and the usual $SU(2)$ link variables, $U(b)$.
The key is to identify a specific representative which  gives a correspondence between the two configurations.  As a 
consequence, the operators specific to the Tomboulis configurations can be measured on $SU(2)$ configurations as 
we show in Section 3.

In the Tomboulis formulation, negative values of  $\eta(p) \sigma(p)$ are constrained to form co-closed
vortex sheets that we denote as $\sigma - \eta$ vortices.  In general, the two species form 
patches denoted by $\sigma$ and $\eta$.   Although pure $\sigma$  vortices are ruled
 out as contributors to the continuum string tension,  the disordering signal from the $\sigma$
patches alone give a string tension that {\em increases} in approach to the continuum limit\cite{ah1,ah2}.  
In Section 4 we give numerical confirmation of this and resolve the connection between this
result and  known scaling results.

These results allow for a direct comparison of these gauge invariant vortex counters with 
those defined  by the projection vortex algorithm\cite{dfgo}.  It is also interesting to note that 
in a particular representative the projection vortices themselves are present in the Tomboulis formulation.
We describe this in Section 5.   In Section 6 we look for coincidences between these two 
types of vortices.

\section{$SU(2)$ configurations in $SO(3) \times Z(2)$ variables} 
In ref\cite{ah1} we rederived the Tomboulis  $SO(3) \times Z(2)$ form\cite{kt1} for the 
$SU(2)$  partition function.  Let us consider only periodic boundary conditions, eliminating
complications due to a twist.
\be
Z  
= 
\sum_{\sigma(p)}
\int 
\left[
    d U(b)
\right]
\prod_{c}
 \delta\left(\sigma(\partial c) \eta(\partial c)\right)
\exp
\left(
   \beta 
   \sum_{p} 
   \half|\lab{tr} [  U(\partial p)]| 
   \sigma(p)
\right),
\label{z}
\ee
where 
\be
\lab{tr} [  U(\partial p)] =  \eta(p) | \lab{tr} [  U(\partial p)] |.
\ee
The $\sigma(p)$
are a new set of independent variables living on plaquettes. 
The $m  \times n$ Wilson loop in these variables  includes tiling factors $\sigma(p)$ and $\eta(p)$,
\be
W_{m \times n} &=& \lab{tr} [  U(C)]  \eta(S) \sigma(S)|_{C = \partial S}, 
\label{wilson}
\\ \nonumber \\
P =  W_{1 \times 1} &=&  \lab{tr} [  U(\partial p)] \eta(p)  \sigma(p) 
= |\lab{tr} [  U(\partial p)] |  \sigma(p),
\label{plaq}
\ee
where $S$ is any surface that spans $C$.  Note that $\sigma(p)$ gives the sign of the plaquette.
The delta function constraint
enforces an even number of negative $\sigma(p)$ or $\eta(p)$ faces on all cubes.  As a consequence, the
elementary excitations of these Z(2) valued variables taken together from co-closed vortex sheets. 
We denote these in general as $\sigma - \eta$ vortices which in the degenerate cases are pure  $\sigma $  or 
pure $\eta$ vortices. 
The partition function, Eqn.(\ref{z}), is invariant under sign flips of the links. 
On a lattice with $n$ links 
there are $2^n$  {\em representatives} of the $SO(3)$ symmetry obtained by flipping the signs of 
the links.
 Such a transformation flips
 the sign of the 6 $\eta(p)$'s forming the co-boundary of a link.  This in turn will result in
$0$ or $2$ sign flips for all cubes, leaving the constraint satisfied.  This process 
will either create an elementary vortex of negative $\eta(p)$'s or deform 
an existing vortex.   The plaquette and all Wilson loops are invariant under this operation.

One can reach all configurations through local updates of the independent variables $ U(b)$
and $\sigma(p)$ while maintaining the cube constraint\cite{ah1}.   Starting from a configuration in which
the cube constraint is satisfied,  e.g. a cold configuration, (i) one first updates the links $U(b)$.  
If  the sign of $ \lab{tr} [U(\partial p)] $ in the co-boundary of the link flips then  $\eta(p)$ 
also flips.  The cube constraint will be maintained if one also flips the sign of the corresponding 
$\sigma(p)$.  Second, (ii), flip the sign of all six $\sigma(p)$'s forming the co-boundary of each 
link.  This will also maintain the constraint.

We now want to relate these configurations to SU(2).  Using the freedom of 
choosing a representative we can modify the algorithm.  Note that in step (i)
a negative $\eta(p)$ is always accompanied by a corresponding negative $\sigma(p)$.
We can modify step (ii) so that this property holds there too.  This is achieved by
flipping the sign of the link $U(b)$ that defines the six co-boundary plaquettes. 
This would flip the signs of the corresponding six $\eta(p)$'s.

This particular choice of representative maps the $SO(3) \times Z(2)$ configurations to 
$SU(2)$.   Note that in this representative, denoted $\widetilde{U}(b)$, 
$\eta(p)\sigma(p) = 1$, and the cube constraint, Eqn.(\ref{z}),
is automatically satisfied. Hence  $Z$ simplifies to
\be
\widetilde{Z}
&=& 
\int 
\left[
    d \widetilde{U}(b)
\right]
\exp
\left(
   \beta 
   \sum_{p} 
   \half\lab{tr} [  \widetilde{U}(\partial p)] 
\right),
\ee
The Wilson loops also  simplify
\ben
W_{m \times n} &=& \lab{tr} [  \widetilde{U}(C)].
\een
Further, this particular representative is characterized by the fact that there are no $\sigma-\eta$
vortices since negative $\sigma$'s are paired with  negative $\eta$'s.
\be
\widetilde{U}(b) \;\;\;\; \Longleftrightarrow \;\;\;\; U^{SU(2)}(b)
\;\;\;\; \Longleftrightarrow 
\;\;\;\; (\sigma - \eta) \lab{ vortices absent}
\label{tilde}
\ee

The $SU(2)$ and $SO(3) \times Z(2)$ local update algorithms are known independently to be ergodic,  and since
there is a correspondence between the two in a particular representative, then the difference between
the two link configurations, simply reflects the difference in representatives.

From a practical point of view this means that we can use the less cumbersome  $SU(2)$ updates
to produce configurations and then use the $SO(3) \times Z(2)$ formalism to 
explore representatives and the corresponding vortex structure.

\section{Kovacs-Tomboulis vortex counters in $SU(2)$}

We have seen in the previous section how to generate the $SO(3) \times Z(2)$ configurations using the 
regular $SU(2)$ algorithm. We show here how to measure 
the operators defined by Kovacs and Tomboulis\cite{kt1} 
on the equivalent $SU(2)$ configurations.

First, let us review these definitions. 
 The sign of the Wilson loop is given by three components: a $\sigma$ tiling, an $\eta$ tiling and the
sign of the contour. They defined three vortex counters describing the number 
of vortices $mod(2)$ linking a particular Wilson loop
\ben
N_{thin}(S)&=&\sigma(S), \;\;\;\;\;\;\;\;\; \sigma(S) = \prod_{p\in S} \sigma(p), \nonumber\\
N_{thick}(S)&=&\lab{sign}\,\, \lab{tr} [ U(C) ]\,\, \eta(S), 
 \;\;\;\;\;\;\;\;\; \eta(S) = \prod_{p\in S} \eta(p),\nonumber \\
N_{hybrid}&=&N_{thin}(S)\times N_{thick}(S)= \lab{sign}\,\, W_{m\times n}.
\een
The first two counters depend on the surface chosen to tile the Wilson loop. The hybrid counter is just
the product of the first two counters and it is independent of the surface $S$. Thus $N_{thin}$ and 
$N_{thick}$ change sign simultaneously as we change the surface $S$. 
If $N_{thin}$ or $N_{thick}$ $= -1$ for all surfaces tiling the Wilson loop then we have
a pure {\em thin} or {\em thick} vortex linking with the Wilson loop (or an odd number of them). If they change 
value as we change $S$ but $N_{hybrid}=-1$ then we have a {\em hybrid} vortex (or an odd number of them)
linking the Wilson loop.

This  set of operators can be defined on  $SU(2)$ configurations.  
Following the same argument as above we take a $SO(3) \times Z(2)$ configuration 
in the particular representative that has no $\sigma - \eta$ vortices, denoted $\widetilde{U}(b)$ in Eqn(\ref{tilde}).
Since for every plaquette with $\sigma(p)=-1$ we also have 
$\eta(p)=\lab{tr} [\partial \widetilde{U}(p)]=-1$, 
the $SU(2)$ operator for $\sigma$ is
$$
\sigma^{SU(2)}(p)=\lab{sign} \,\,\lab{tr} [\partial \widetilde{U}(p)].
$$
Using this we can define the thin vortex counter on $SU(2)$ configurations to be
$$
N_{thin}^{SU(2)}(S)=\prod_{p\in S} \lab{sign} \,\,\lab{tr} [\partial \widetilde{U}(p)].
$$
We have already noted that the hybrid counter is  the sign of the Wilson loop and thus we have
$$
N_{hybrid}^{SU(2)}=\lab{sign} \,\, \lab{tr} [\widetilde{U}(C)].
$$
To define the thick counter  we use the fact that $N_{thick}(S) = N_{thin}(S)\times N_{hybrid}$,
$$
N_{thick}^{SU(2)}(S)=\prod_{p\in S} \lab{sign} \,\,\lab{tr} [\partial U(p)]\times \lab{sign} \,\, \lab{tr} [U(C)].
$$

Although the presentation here was
more intuitive than rigorous it is easy to show that these operators are indeed equivalent
 in the sense that they have the same average on
$SU(2)$ configurations as the original vortex counters on $SO(3)\times Z(2)$ configurations
and similarly for polynomial functions.

We can measure these operators using standard  $SU(2)$ simulations.
In the next section we investigate the
interquark potential based on these counters. 
Furthermore, we can now compare these vortices with P vortices on the same configuration which we do in 
the last two sections.

\section{Vortex Potentials}

Although it is straightforward to measure  $N_{thin}$,  $N_{thick}$ and  $N_{hybrid}$, it is highly
problematic to correlate these measurements with linkings of the corresponding species of
vortex. This is because of the proviso that the value be independent of surface. 
It is very time consuming to generate all possible surfaces that tile a particular Wilson loop.
Moreover, in a configuration that has {\em monopoles} (cubes where the product of $\sigma$ on all six face is
negative) these pure vortex counters are difficult to define.  According to the above
definition all vortices are hybrid.  Due to these problems we restrict our attention to only the minimal 
surface to define the vortex counters with the understanding that it is ambiguous. 
If, for example, the  {\em thin}  vortex counter $= -1$ for the minimal surface
then we detect either a trapped  {\em thin} or {\em hybrid} vortex.   We expect that
 averages of these operators are independent of surface.

To extract the vortex potentials we measured the Wilson loops and vortex counters at 
$\beta=2.3$,  $2.4$  and $2.5$ on a $22\times 14^3$ lattice, using $3000$, $1000$ and $1228$ 
configurations respectively.
We thermalized using $1000$ updates and the measurements were 
separated by $40$ updates. The acceptance was calibrated to be approximately $50\%$.

The contribution to the potential from the three types of vortex counters is
$$
V_{counter}(R)=-\lim_{T\pe\infty} \frac{1}{T} \ln \la N_{counter}(W(R, T)) \ra,
$$
where $N_{counter}(W(R,T))$ is the counter signal for that particular Wilson loop (taking values  $\pm 1$). 
To determine the potential for a particular 
$R$ we use Wilson loops $W(R, T)$ and an array of $T$'s that are large enough for the 
exponential behavior to set in and do a fit with an exponential in $T$.

We extracted the string tension, $\sigma$, and checked scaling by fitting these data with the function
$$
V(r)=\sigma r - \frac{e}{r}+V_0,
$$
where $e/r$ represents the Coulomb part of the 
potential at short distances and $V_0$ is a self-energy. The parameters $\sigma$ and $e$ are expected 
to scale whereas $V_0$ which depends strongly on the cut-off is not expected to scale. 
Using the physical value of the string tension $\sigma=(440 MeV)^2$ we determine the 
lattice spacing. The results are in the Table \ref{tableTom1}. 

\begin{table}[ht]
\begin{center}
\caption{String tension and lattice spacing. \label{tableTom1}}
\begin{tabular}{|c|c|c|c|c|}
\hline
$\beta$ & $\sigma$ [lattice units]& lattice spacing [fm] & e [natural units] & $V_0$ [GeV]\\ \hline\hline
2.3 & 0.157(14) & 0.177(8) & 0.193(25) & 0.51(11) \\ \hline
2.4 & 0.083(10) & 0.129(8) & 0.217(24) & 0.78(20) \\ \hline
2.5 & 0.043(1) & 0.093(2)  & 0.211(3)  & 1.07(6) \\ \hline
\end{tabular}
\end{center}
\end{table}

\begin{figure}[p]
\begin{center}
\epsfig{file=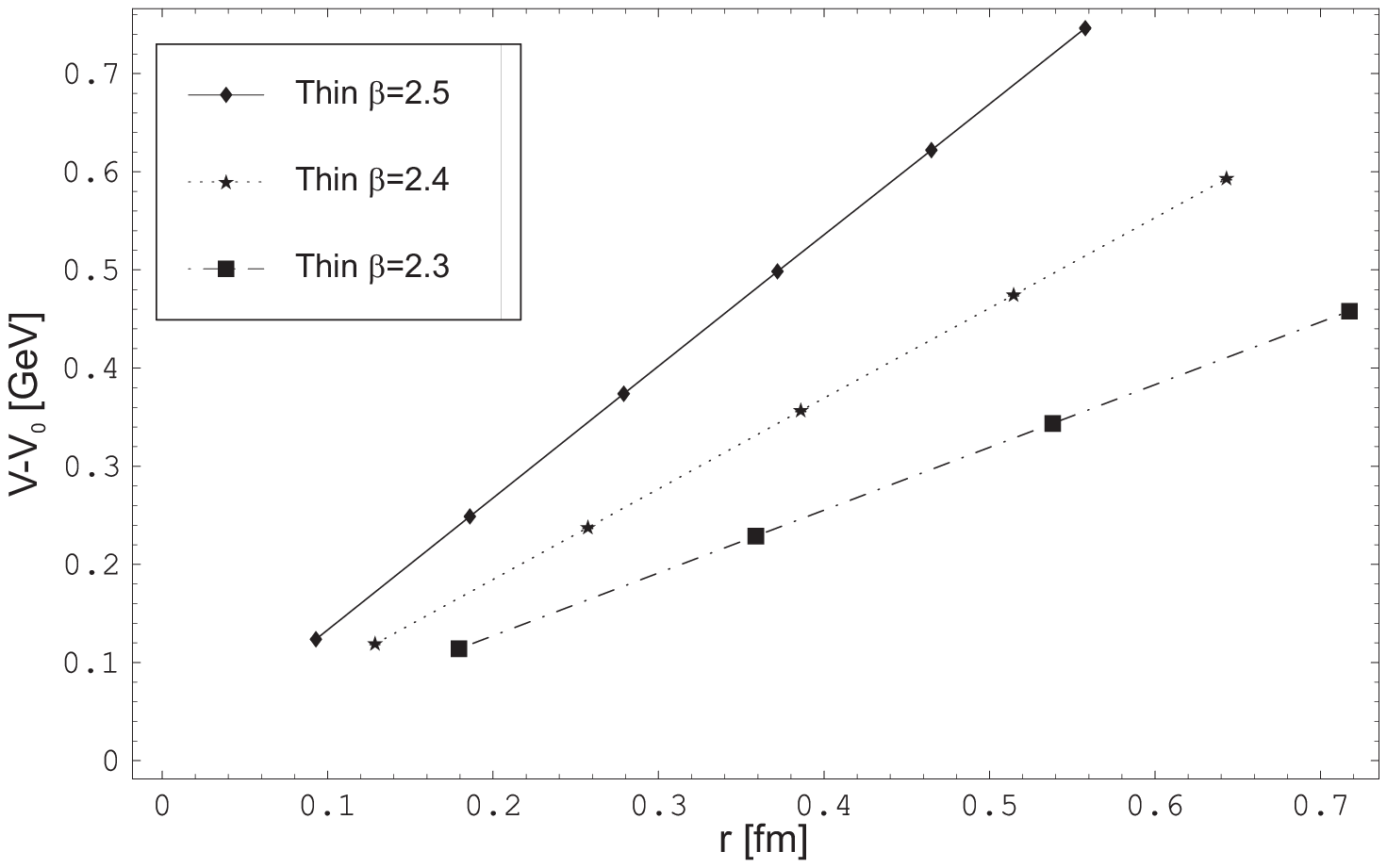, width=10cm}
\epsfig{file=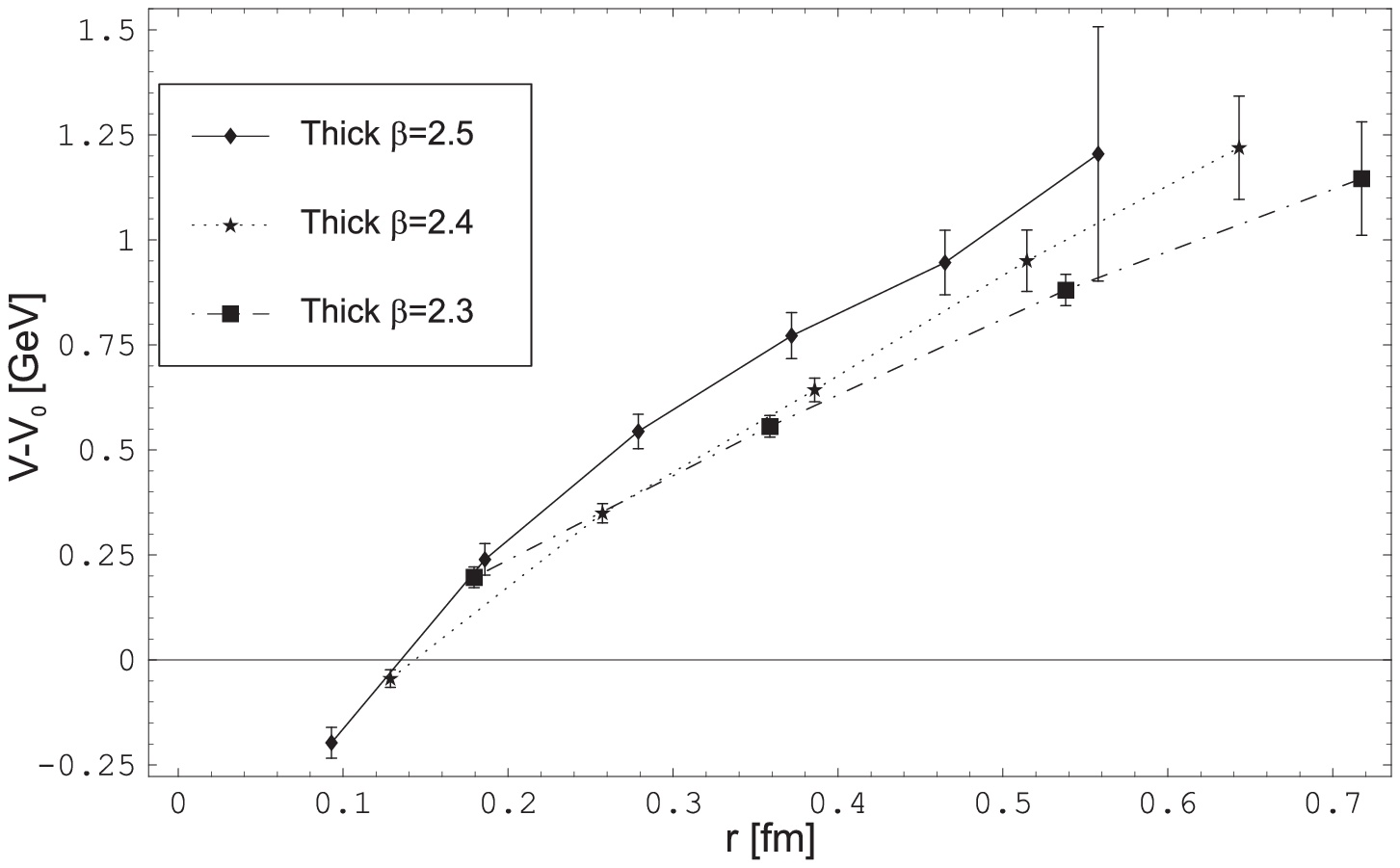, width=10cm}
\epsfig{file=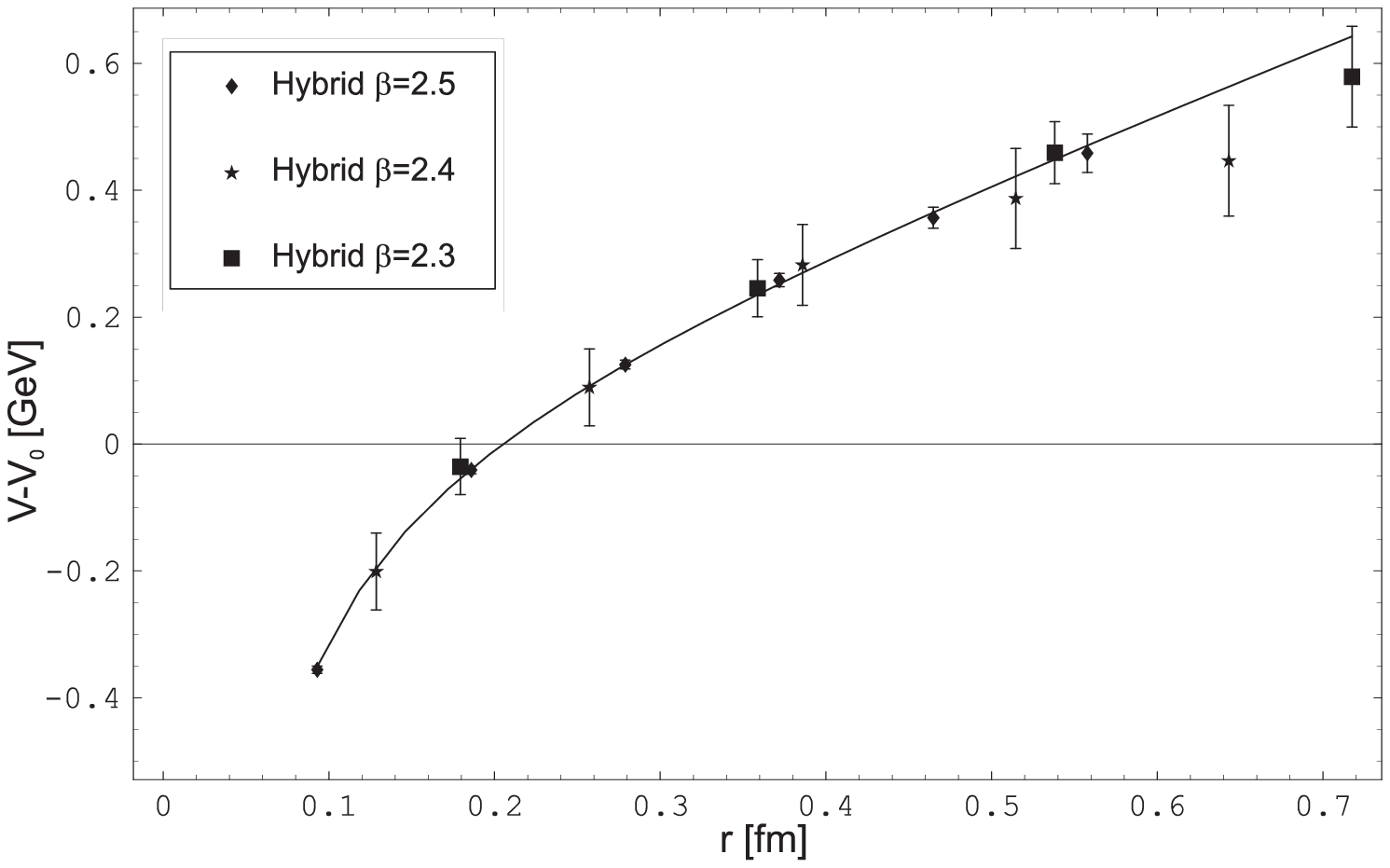, width=10cm}
\end{center}
\caption{Vortex potentials in physical units: thin (top), thick (middle), hybrid (bottom). 
The line in the hybrid plot represents a fit.\label{figTom5}}
\end{figure}

In Fig. \ref{figTom5} we present the scaling graphs for the potentials extracted using the vortex counters. We see
that the hybrid potential scales since it follows exactly the full potential. 
This was first noticed by Kovacs and Tomboulis \cite{kt1}.

We also see that the potential generated by the thick counter differs substantially from the full potential. This
is contrary to our expectations since we expected that the thick counter would show a  behavior  very much
like the hybrid counter. We show below that this is a consequence 
of the existence of thin patches in the hybrid vortices. 

The thin counter produces a very clear string tension. This agrees with our observation \cite{ah1,ah2} of
the string tension in the Wilson loop tagged by thick vortices (the string tension there was roughly $1.1$ where
the string tension at $\beta=2.3$ extracted from thin potential is $1.043(1)$). 

However note that the thin and thick potentials do not scale. Moreover, we see from these plots very clearly 
that the thin potential string tension increases in physical units rather than vanishing\cite{ah1,ah2}. 
Thus the potential produced by thin patches, although not 
relevant for confinement, cannot be disregarded. The fact that the string tension due to the thin patches does not
vanish as we approach the continuum limit is also producing a non-scaling behavior for the thick potential. To see 
this we write down the hybrid counter:
$$
N_{hybrid}=N_{thin}\times N_{thick}
$$
If the thin and thick counters were completely uncorrelated then we would expect that:
$$
\la N_{hybrid}\ra = \la N_{thin} \ra \times \la N_{thick}\ra
$$
Since we know now that $\la N_{thin}\ra\sim e^{-\sigma A}$, with increasing $\sigma$ (in physical units) as we 
approach the continuum limit, the hybrid counter would also have a non-physical string tension. However, we know 
that the hybrid potential scales properly (since it behaves exactly like the full potential) and thus the thin 
and thick counters cannot be uncorrelated\cite{ah2}. 

The correlation comes from the hybrid vortices since our counters 
measure the signal only on a single surface. A hybrid vortex produces a thin or thick signal 
depending on how it pierces the minimal surface. Thus both the thin  and  thick counters includes 
extraneous signals due to hybrid vortices. Since we believe that the pure thin vortices cannot produce any string 
tension as we approach the continuum limit we assume that the string tension that we see in the thin 
counters is due to these hybrid vortices (more precisely  the thin patches in the hybrid vortices). 

\begin{figure}[p]
\begin{center}
\epsfig{file=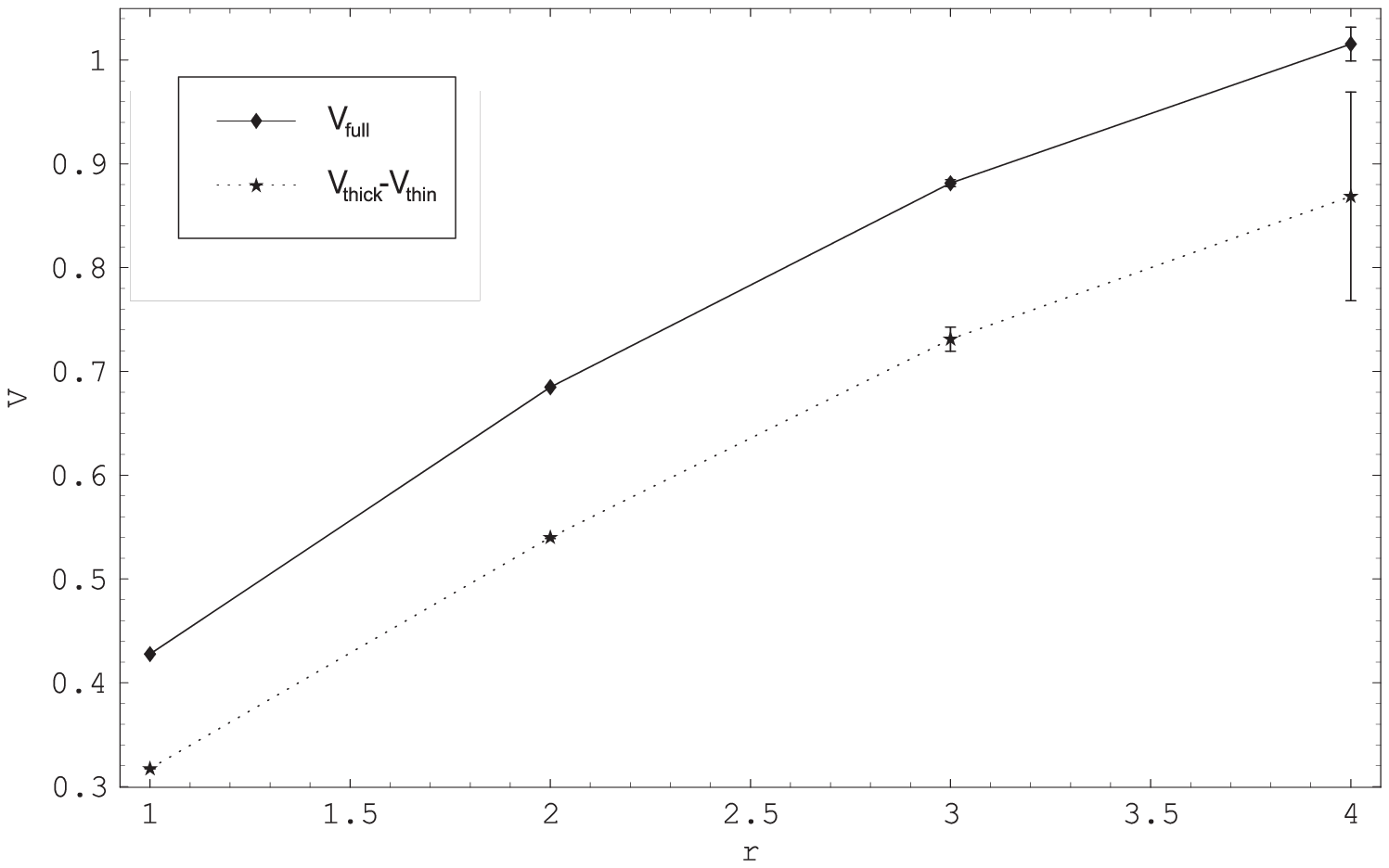, width=10cm}
\epsfig{file=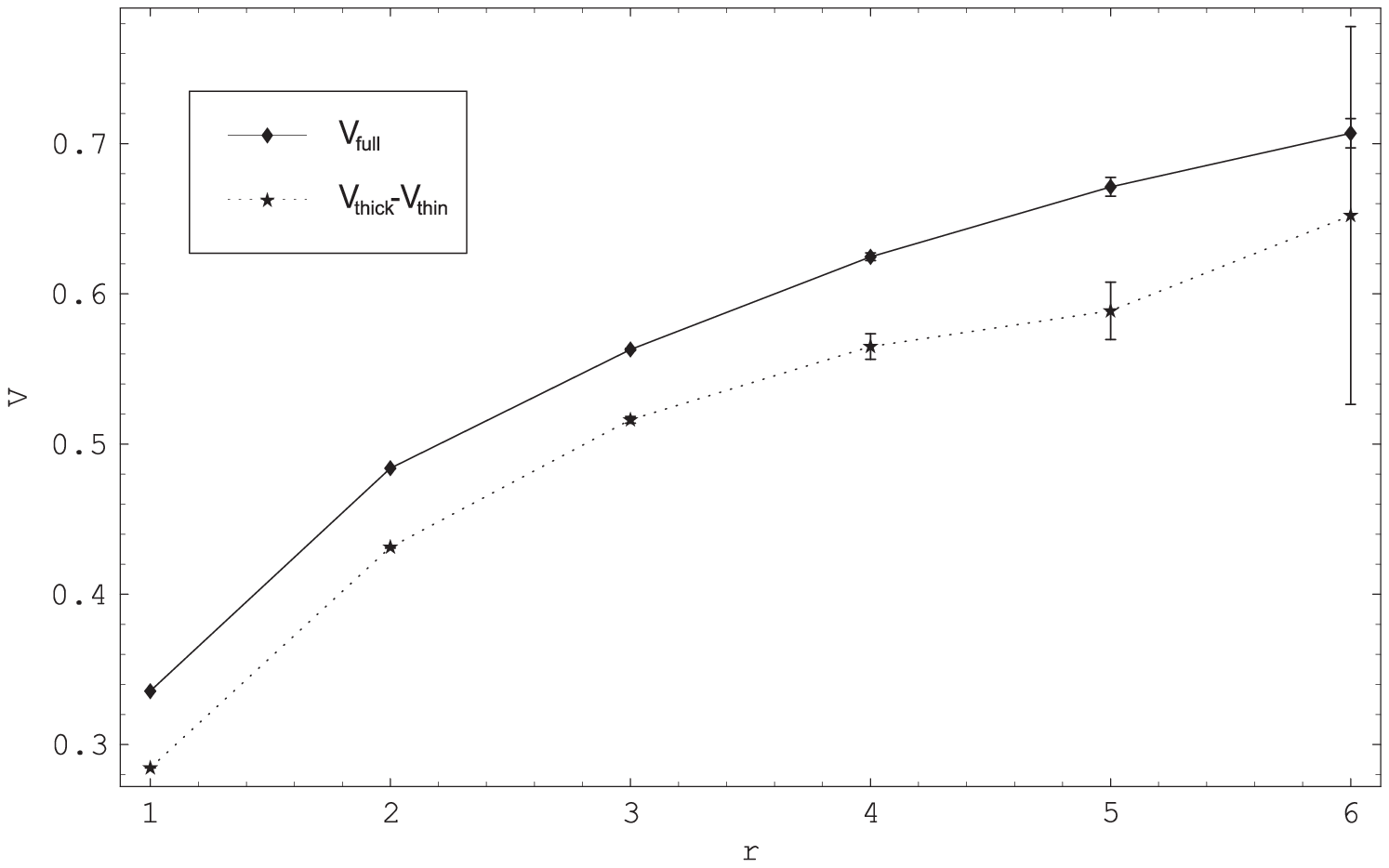, width=10cm}
\epsfig{file=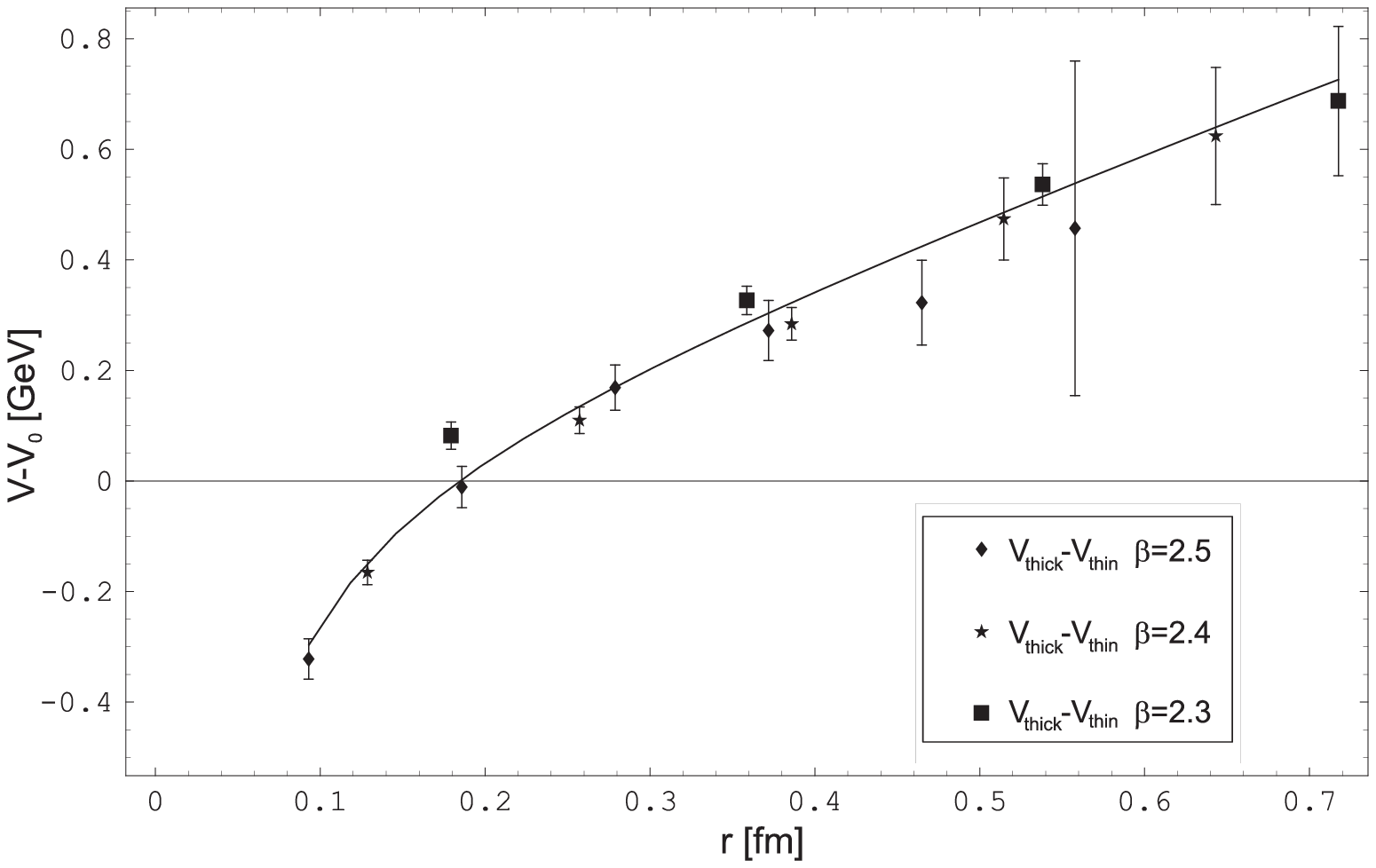, width=10cm}
\end{center}
\caption{The difference between the thick and the thin potential plotted along with 
the full potential. The graphs at the top and middle are for $\beta=2.3$ and $\beta=2.5$ 
respectively.  The bottom is both data sets on a scaling  graph.\label{figTom6}}
\end{figure}

In order to see the properties of the pure thick vortices we need  to remove the contribution due to the 
thin patches of hybrid vortices.  One way to do it, if the above reasoning is true, 
is by subtracting the thin potential from the 
thick potential. In Fig. \ref{figTom6} we plot this difference. We see that apart from a constant the 
full potential and difference of the thick and thin potentials are the same. The string tension recovered from the 
scaling graph is the same within the error bars with the full string tension.

\section{Projection Vortices}

To define projection vortices, (P vortices) \cite{dfgo}, one fixes
to a standard gauge, e.g. the maximal center gauge.
Let us set aside the  specifics of gauge fixing.

There are two distinct ways to identify P vortices in $SO(3) \times Z(2) $ configurations
giving the same result.

(i)  After gauge fixing, one projects the full links $U(b)$ 
onto $Z(2)$ link variables $u(b)$, taking the values  
\ben
sign \;\; \lab{tr} [U(b)] \rightarrow u(b). 
\een
The negative plaquettes in this $Z(2)$ theory form co-closed surfaces, i.e. giving thin vortices. These
would be the P vortices in an SU(2) configuration.  But that is not the  case here.

The problem with this procedure is apparent in noting that changing representatives 
changes the locations of these vortices.  The action is invariant under these transformations and hence 
the locations of these thin vortices are not uniquely defined.    

In order to get P vortices we need to 
include $\eta(p)$ and $\sigma(p)$ tiling factors just as
in the case of full Wilson loops, Eqn.(\ref{wilson}).   
Therefore the P vortices are given by the occurance of negative values of $u(p)$ where
\be
u(p)  =  u(\partial p) \eta(p)  \sigma(p) 
\label{plaq2}
\ee
where $\eta(p)$ and $\sigma(p)$ are defined in the unprojected full theory.  
This definition is representative independent in the same sense that the 
Wilson loop is in the full theory.

(ii) There is a simpler identification of P vortices that makes use of the freedom to 
choose a particular representative which we denote by $\widehat{U}(b)$.
Starting with a given $SO(3) \times Z(2) $ configuration let us make a single sweep flipping the signs
so that for all links
\ben
\lab{tr} [\widehat{U}(b)] \ge 0.  
\een
In this representative the projected links, $u(b)$, are all positive and there are no negative 
values of $ u(\partial p)$ and no vortices due to the $u(b)$ variables. 
But that means that the tiling factors themselves give the P vortices. 
The $\sigma  - \eta$ vortices are already the $P$ vortices themselves in this particular representative.  
\ben
\widehat{U}(b) \;\;\;\; \Longleftrightarrow \;\;\;\; 
\;\;\;\; (\sigma - \eta) \lab{ vortices are the same as  P vortices }
\een

Although this gives the P vortices, there is as yet no approximation.  One has transferred the 
sign from one factor to another  making up the Wilson loop, Eqn.(\ref{wilson}).

The approximation comes when one replaces
\be
\lab{tr}   [\widehat{U}(C)] \rightarrow  +1.
\label{projapprox}
\ee
The success of this approximation is dependent on the success of the gauge fixing algorithm 
in transferring the confinement physics from the contour to the tiling factors.

\section{Vortex Comparison}

\begin{figure}[p]
\begin{center}
\epsfig{file=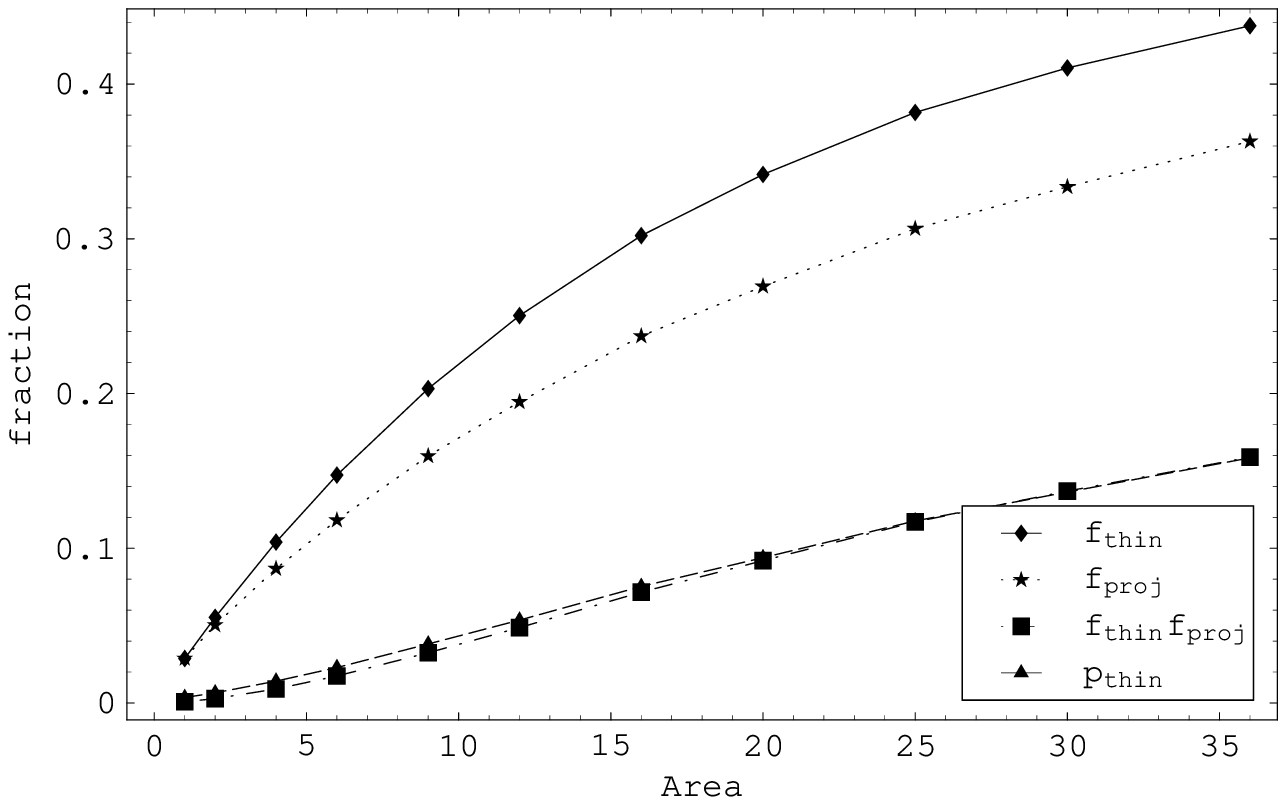, width=10cm}
\epsfig{file=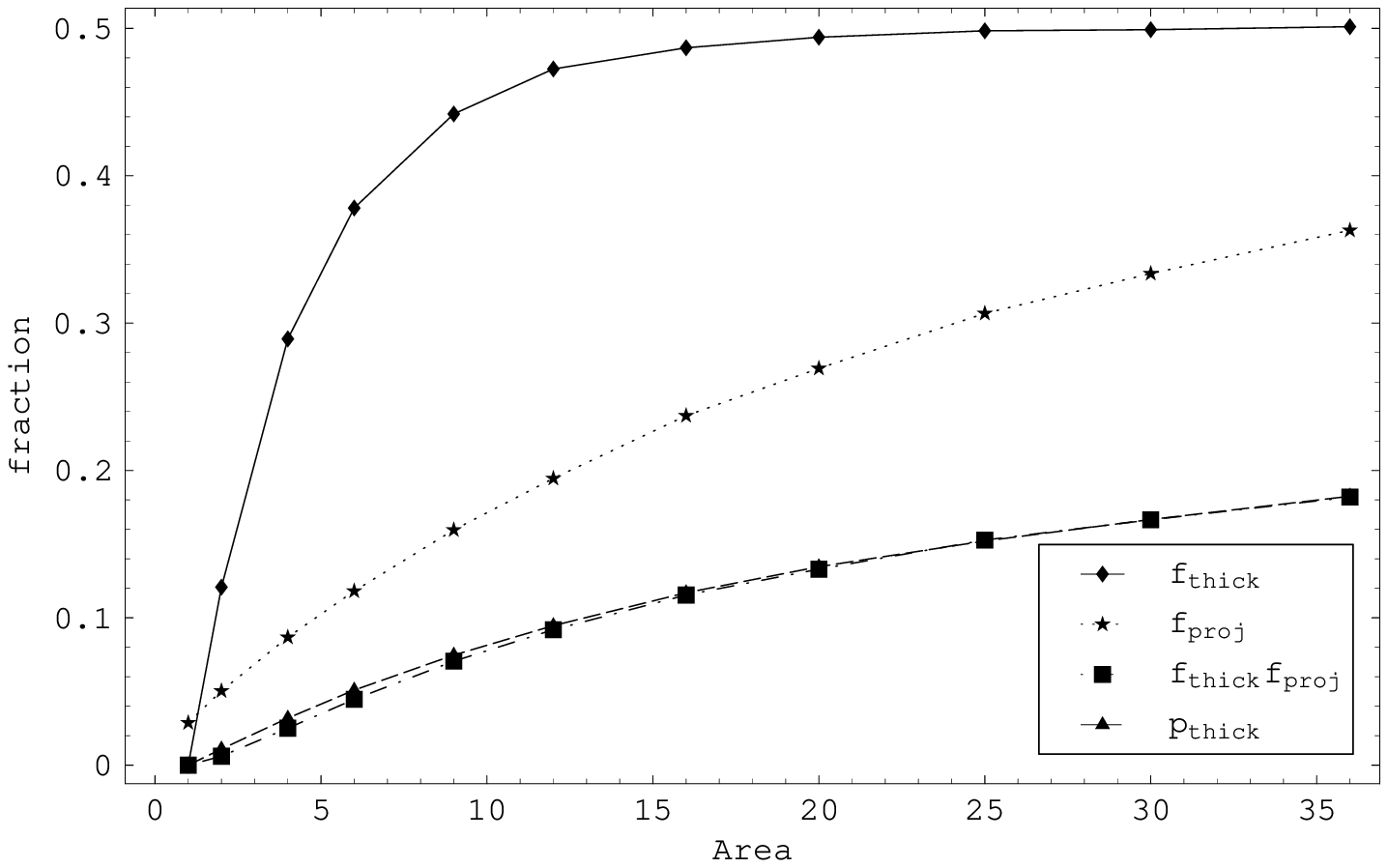, width=10cm}
\epsfig{file=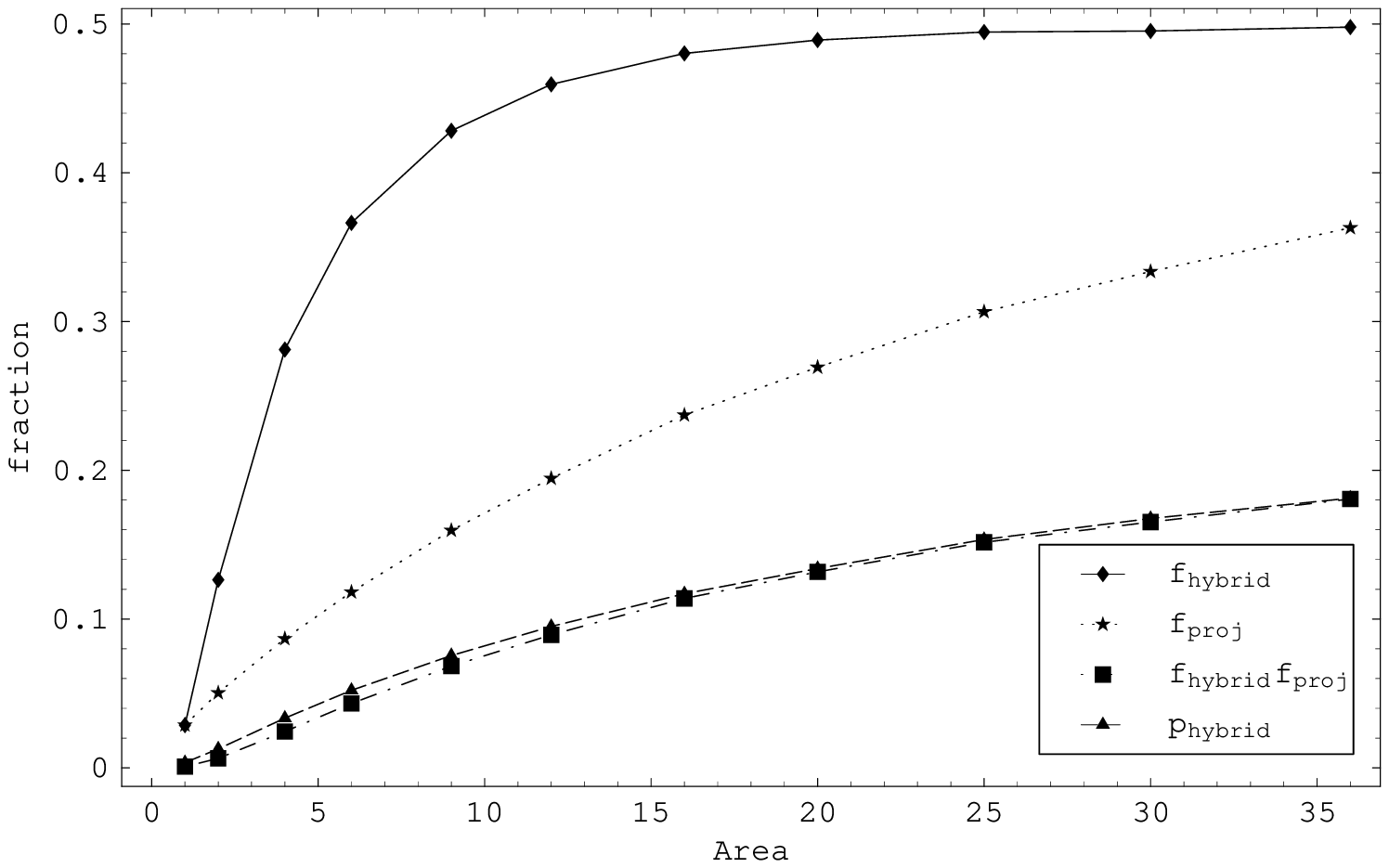, width=10cm}
\end{center}
\caption{K-T counters and their coincidence with the P vortices. 
The $p_{counter}$ is very close to the lower bound $f_{proj}\times f_{counter}$ 
which indicates no correlation.\label{figTom7}}
\end{figure}

The K-T definition for vortices\cite{kt1} is appealing since it is  gauge invariant 
 but they are hard to localize on a lattice. 
P vortices \cite{dfgo}, on the other hand, are easy to localize but are not gauge invariant. 
It is interesting to see if these two definitions agree. 
We now have the tools to compare these definitions of  vortex counters 
on the same configurations. 

Our first test was to take a thermalized $SU(2)$ gauge configuration and project it.  
We fixed to the direct center gauge\cite{dfgo} without preconditioning\cite{kt3} or simulated 
annealing\cite{b}
where the evidence for P vortex dominance is strongest.   We
measured  the K-T counters on the original gauge configuration and the P vortex counter 
on the projected configuration. We counted only the fraction of loops of a certain size that 
produces a negative signal,
$$
f_{counter}=\la \frac{1}{2} (1-N_{counter})\ra,
$$
where the counter can be thin, thick, hybrid or projection. We then measured the coincidence 
between the P vortex counter and one of the K-T counters. This  measures the fraction 
of Wilson loops of a certain size that has both the P vortex counter and that particular K-T 
counter negative,
$$
p_{counter}=\la \half (1-N_{projection}) \times \half (1-N_{counter})\ra.
$$
where the counter can be thin, thick or hybrid. If the vortices are completely uncorrelated then,
$$
p_{counter}=\la \half (1-N_{projection})\ra \times \la \half (1-N_{counter})\ra=f_{projection}\times f_{counter}.
$$
If they are completely correlated then,
$$
p_{counter}=\min \{f_{projection}, f_{counter}\}.
$$
These are the bounds on the coincidence counter. 
If the $p_{counter}$ approaches the lower bound $f_{proj}\times f_{counter}$ then 
the vortex counters are uncorrelated and we conclude that the physics that generates 
the counters is different. If the coincidence counter is closer to the upper bound then we 
conclude that the counters detect the same structures.

\begin{figure}[p]
\begin{center}
\epsfig{file=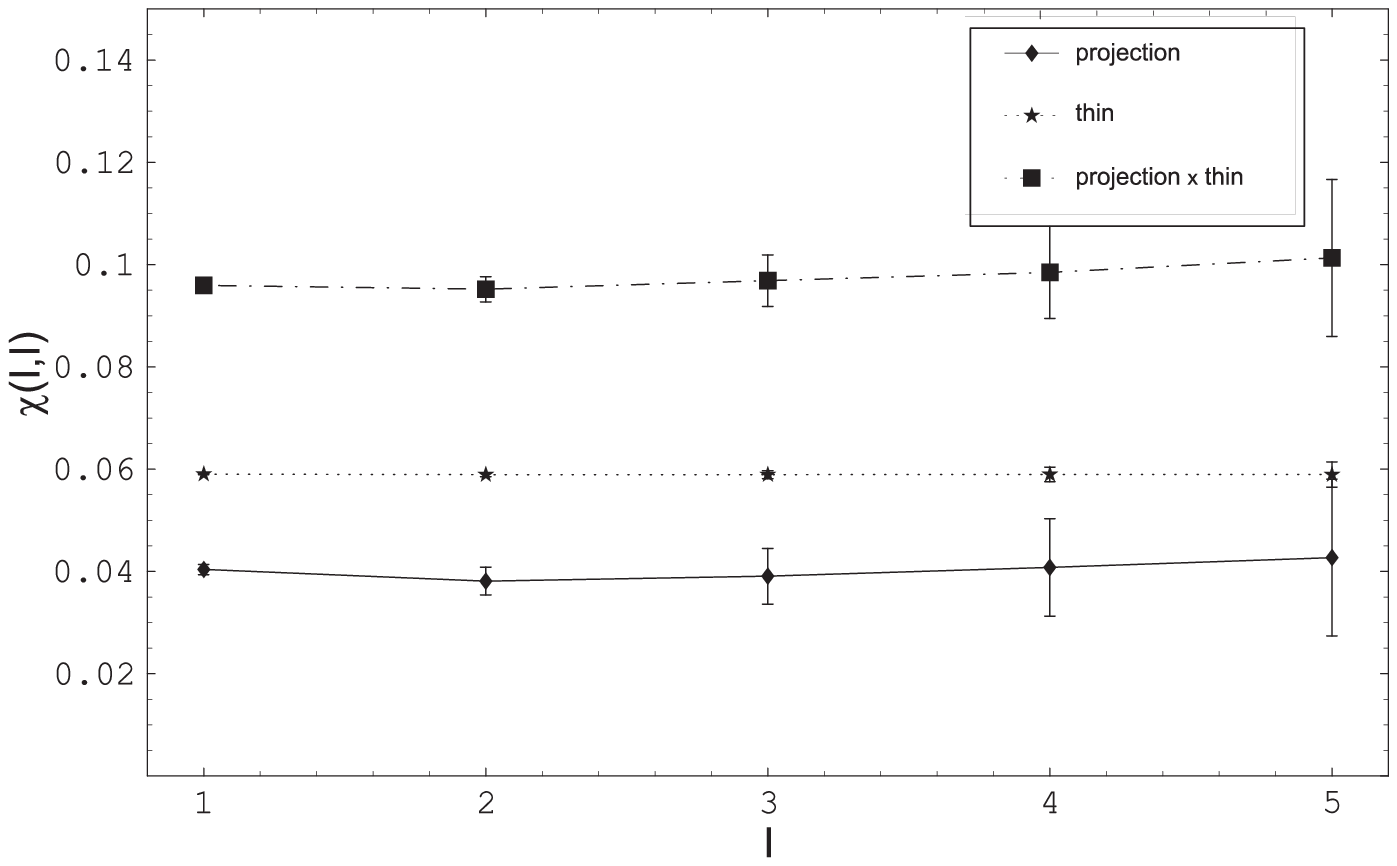, width=10cm}
\epsfig{file=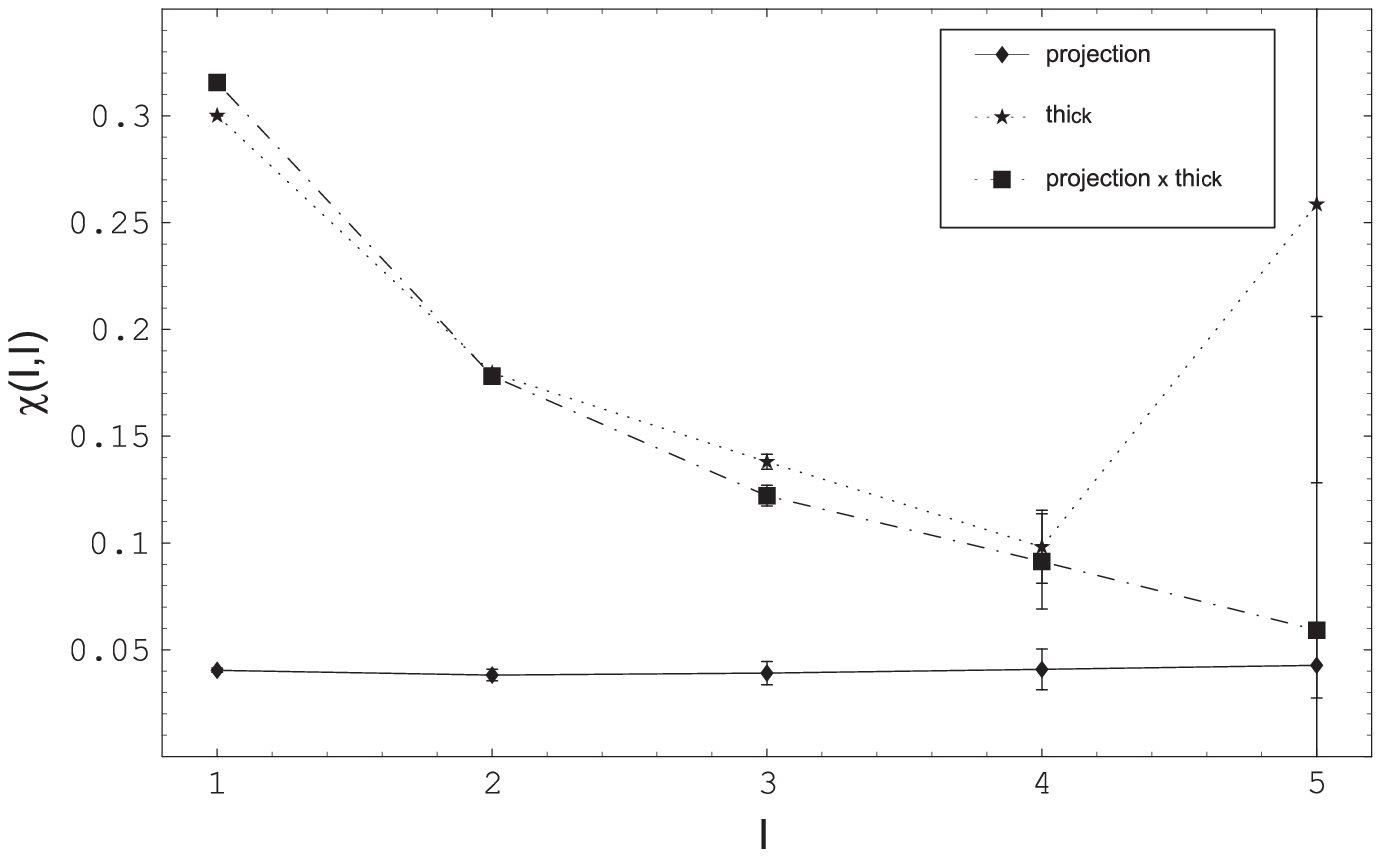, width=10cm}
\epsfig{file=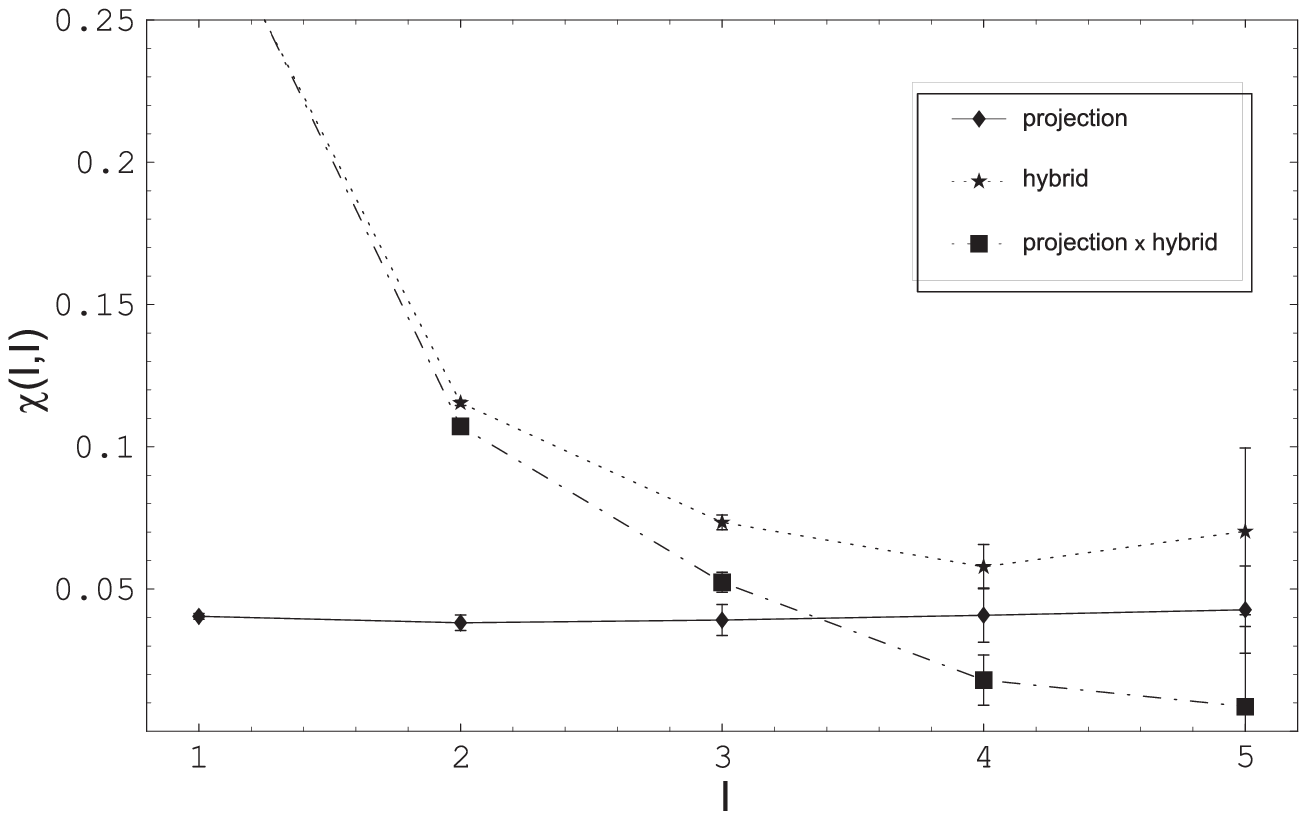, width=10cm}
\end{center}
\caption{Creutz ratios for the products between the  P vortex counter and K-T counters.\label{figCreutz}}
\end{figure}

The results are presented in Fig. \ref{figTom7}. 
We see that the counters show no correlation. 
This indicates  that  P vortices and K-T vortices are different objects.

However, there is another possibility which leads to our second test. 
The projection procedure can produce different results for gauge 
equivalent configurations. The argument is that when we have a thick center vortex the projection 
produces a P vortex somewhere inside the core. However, this P vortex can
be anywhere inside the thick core depending on the gauge copy used. Thus 
the correlation signal might be lost due to P vortices fluctuating in an out of the perimeter.

To determine whether the perimeter produces the decorrelation we employed a method 
used to determine the self-correlation of P vortices \cite{dfggo}. We looked at products
of the P vortex  and K-T vortex counters,
$$
\la N_{counter}(W_{m\times n}) N_{projection}(W_{m\times n})\ra.
$$
Each of the counters produces a string tension. If they are uncorrelated their product will 
exhibit a string tension equal to the sum of their string tensions. On the other hand, if they
are correlated and identify the same vortices, and if the vortices they identify are indeed 
responsible for confinement then the product may have a perimeter behavior but 
will have no string tension.

In Fig. \ref{figCreutz} we present the Creutz ratio for the products between the P vortices and
 Tomboulis vortices together with the Creutz ratios for each counter alone. Our simulation was
run on a $16^4$ lattice at $\beta=2.5$. We used $1000$ sweeps to thermalize the lattice and
we made $1000$ measurements separated by $40$ updates.

We see from these graphs that the thin vortex counter shows no correlation with the 
P vortex counter. However, both thick and hybrid counters are showing a strong correlation
with the P vortex counter. 

In the case of the hybrid vortex counter we expected the 
correlation since we know that by eliminating P vortices the full Wilson loop looses the string
tension \cite{dfgo,dfggo} and that the hybrid counter produces the same potential
as the full Wilson loop \cite{kt1}. The thick vortex correlation with the P vortex is 
due to the pure thick part of the counter. We know that the thick counter also includes a part
due to thin patches and thus we don't expect the product of P vortex counter and thick
counter to lose the string tension since we know that the thin vortex counter is uncorrelated
with the P vortices. Our expectation is that this product will exhibit a string tension that
is equal with the one generated by the product of thin vortex counter with the P vortex counter.
Unfortunately, the precision of our data doesn't allow us to check if this is indeed true.

\section{Summary and Conclusions}

We have used the freedom of picking a representative to uncover some kinematical relationships.
On the one hand we chose a representative, denoted $\widetilde{U}(b)$ which eliminated $\sigma - \eta$ vortices allowing
the mapping of the $SO(3) \times Z(2)$ configuration space onto $SU(2)$.  And second by choosing another
representative in which $\lab{tr}  [\widehat{U}(b)] \ge 0$ for all links we found the $\sigma - \eta$ vortices 
to be identical to P vortices.  The {\em projection approximation} has the added 
steps of gauge fixing and,  after picking the representative, 
setting the contour integral $\lab{tr} [\widehat{U}(C)] = 1$.  

It is interesting to note that  it is not necessary to define an underlying $Z(2)$ gauge 
theory to identify $P$ vortices.   
These structures are already present in the full $SO(3) \times Z(2) $ 
version of the theory.

We find it surprising that  the string tension in the ``thin potential" doesn't vanish in the continuum limit. 
On the contrary it becomes larger. We 
argued that this must be due to thin patches of hybrid vortices.  The counters measured on a
single surfaces can not distinguish between the two.   

We further see that there is a strong correlation between the thick and thin counters.  
The thick counter on a single surface can not distinguish thick vortices from thick patches of 
hybrid vortices.  Hence as a corollary, we expect this anomalous string tension in
the ``thick potential" which we also see.   This is further support that hybrid vortices are
responsible.  

By subtracting the ``thin potential" from the ``thick potential" we conjecture that we see the
potential due to thick vortices alone.  To see that this is indeed the potential due to 
pure thick vortices a more careful analysis is required. We need to 
first find a definition for pure thick vortices that works in a general configuration  including 
those that have monopoles. 

An alternative  is to use the definition that we have now but 
generate configurations that have no monopoles. In such configurations we have only pure thin and thick vortices. 
It is very likely that in such an approach that the thick potential will be identical with the full potential (at 
least at large distances) since we expect that the pure thin vortices produce at best a perimeter law. The problem 
with this approach is that we are changing the dynamics of the system by forbidding the monopoles.

\section{Acknowledgments}

We are pleased to thank E. T. Tomboulis, S. Olejnik and S. Cheluvaraja for 
helpful discussions.  
This work was supported in part by United States Department of Energy
grant DE-FG05-91 ER 40617.

\end{document}